\documentclass[%
superscriptaddress,
reprint
floatfix,
amsmath,
amssymb,
pra,
]{revtex4}
\usepackage{boxhandler}
\usepackage{graphicx}
\usepackage{dcolumn}
\usepackage{bm}
\usepackage{epsfig}
\usepackage{color}
\usepackage{longtable}
\usepackage{cancel}
\usepackage{float}
\usepackage{bm}
\usepackage[french,english]{babel}
\usepackage[OT1,T1]{fontenc}
\usepackage[normalem]{ulem}
\usepackage{lscape}

\usepackage{longtable}

\begin{document}

\title{Weak Correlation and Strong Relativistic Effects on the \\ Hyperfine Interaction in Fluorine  }

\author{Fatima Zahra Boualili}
\affiliation{Laboratoire d'\'Electronique Quantique, Facult\'e de Physique, USTHB, BP32,
El-Alia, Algiers, Algeria}

\author{Messaoud Nemouchi}
\affiliation{Laboratoire d'\'Electronique Quantique, Facult\'e de Physique, USTHB, BP32,
El-Alia, Algiers, Algeria}

\author{Michel Godefroid}
\affiliation{Spectroscopy, Quantum Chemistry and Atmospheric Remote Sensing (SQUARES), CP160/09, Universit\'{e} libre de Bruxelles, 1050 Brussels, Belgium}

\author{Per J\"onsson}
\affiliation{Department of Materials Science and Applied Mathematics, Malm\"o University, SE-20506 Malm\"o, Sweden}

\begin{abstract}
     In previous work devoted to {\it ab initio} calculations of hyperfine structure constants in nitrogen and fluorine atoms, we observed sizeable  relativistic effects, a priori unexpected for such light systems, that can even largely dominate over electron correlation. We observed that the atomic wave functions calculated in the Breit-Pauli approximation describe adequately the relevant atomic levels and hyperfine structures, even in cases for which a small relativistic $LS$-term mixing becomes crucial. 
    In the present work we identify new levels  belonging to the spectroscopic terms $2p^4(^3\!P) 3d ~ \; ^{2,4}\!(P,D,F)$ of the fluorine atom, for which correlation effects on the hyperfine structures are  small, but relativistic $LS$-term admixtures are decisive to correctly reproduce the experimental values. The Breit-Pauli analysis of the hyperfine matrix elements nails cases with large cancellation, either between $LS$ pairs for individual hyperfine operators, or between the orbital and the spin-dipole contributions. Multiconfiguration Dirac-Hartree-Fock calculations are performed to support the Breit-Pauli analysis. 
 \end{abstract}

\maketitle

\section{Introduction}\label{sec:intro} 

The development of relativistic theories applied to atoms has greatly contributed to improving the agreement between theory and  observation. Among the methods accounting for relativity we can cite the multiconfigurational Hartree-Fock (MCHF) approach with Breit-Pauli (BP) corrections \cite{Froetal:97b,Froetal:2016a} and the multiconfigurational Dirac-Hartree-Fock (MCDHF) approach with Breit and QED corrections \cite{Gra:94a,Gra:2007a}. The methodological developments, combined with the increasing computer resources, allow for accurate calculations of atomic wave functions, which make it  possible to study rigorously the  balance between electronic correlation  and relativistic effects on atomic properties. ATSP2k~\cite{Froetal:2007a} and GRASP2018 \cite{Froetal:2019a} are codes  built on, respectively, the MCHF+BP  and  MCDHF+Breit+QED approaches. 

 Correlation effects are traditionally  presented as being dominant in light atoms, on the basis of the $Z$-dependent perturbation approach of the non-relativistic Hamiltonian~\cite{Layetal:64a}, while relativistic effects are expected to be more prominent in heavy atoms, due to the large mean velocity of the inner electrons relatively to the speed of light, when increasing the nuclear charge~\cite{Des:83a,Gra:2007a}. 
This  picture is definitely too simple, as explicitly expressed two decades ago by Desclaux's statement~\cite{Des:2002a}: ``It is obvious that correlation and relativistic corrections should be included simultaneously in a coherent scheme.'' It is nowadays acknowledged that relativity has to be taken into account, even for light atoms~\cite{Bieetal:2015a,Bieetal:96a}, to obtain accurate predictions of electronic structures. 

The effects of relativity on the hyperfine interaction in light atoms have been studied in several works \cite{Bieetal:96a,Bieetal:99a,JonBie:2010a,Aouetal:2018a}. In fully relativistic calculations, as in the MCDHF method, the influence of relativity leads to two effects \cite{Des:72a,Pyy:88a}. The first one is a direct effect that results  in the contraction of radial orbitals compared to the nonrelativistic ones. The second one,  an indirect effect, is a consequence of the first,   that manifests itself by an expansion of  radial orbitals. Orbitals characterized by low  angular momentum $l$-values, i.e. $s$ and $p$ electrons, undergo the first contraction effect, while orbitals with larger $l$-values, more efficiently screened due to the relativistic contraction of the $s$  and $p$ shells, are radially outward expanded.  These effects, resulting from the application of purely relativistic methods, have a weak influence on the atomic properties of light elements. In the case of the $1s^2 2s^2 2p^4 3d$ configuration of fluorine $(Z=9)$, the mean radii  
of Dirac-Hartree-Fock (DHF) and Hartree-Fock (HF) orbitals,
$\langle r \rangle_{1s}^{\rm DHF}= 0.17543$ and
$\langle r \rangle_{1s}^{\rm HF}=0.17567$,
 differ relatively by 0.14\%. This contraction effect in fluorine is  rather small in comparison with, e.g. the gold atom, for which the $1s$ orbital  undergoes a relative displacement of the order of 13\%, while the $6s$-contraction is of the order of 17\% , due to the combined direct and indirect  effects of relativity~\cite{Des:73a,Aut:2012a}. 
In the BP approximation, the radial orbitals are frozen from nonrelativistic calculations, while relativity is  captured only through the $LS$-term mixing for a given $J$-value. For light atoms, the inclusion of relativistic effects in the  BP approximation is generally sufficient to  estimate atomic properties accurately. 

Large-scale MCHF calculations combined with non-relativistic configuration interaction calculations  of hyperfine  parameters have been performed successfully in light atoms~\cite{Caretal:92a,JonFro:93a,Jonetal:96a,Jonetal:96b,Godetal:97b}.   In some studies, relativity was included to improve the agreement with observation, either through the Breit-Pauli approximation or using the relativistic interaction configurations (RCI) approach~\cite{Yer:2008a,Jonetal:2010a,CarGod:2011a,CarGod:2011b}. In all these works, the relativistic corrections were not negligible, but remained relatively small, changing the hyperfine parameters by less than a few percent. 
 However, unexpected large deviations have been highlighted in the study of hyperfine structures of some  levels of the fluorine atom, for which the relativistic effects on the hyperfine constants $A_{3 / 2}$ and $A_{5 / 2}$ of the $2p ^ 4 3p \; ^4\!P ^o_{3 / 2,5 / 2}$ levels were estimated to be around 30\%~\cite{Caretal:2013b}. Even larger relativistic effects have been found for other levels~\cite{Aouetal:2018a}, of the order of 35\%  for $A(2p ^ 4 3s \; ^4\!P_{1/2} )$ and,  even more spectacular, reaching 182\% for  $A(2p ^ 4 3p \;  ^4\!S ^o _{3/2} )$.
Aourir {\em et al.}~\cite{Aouetal:2018a} showed that in some cases, although the relativistic effects can be important for the different contributions  to the hyperfine interaction, the global effect of relativity may become relatively small  due to large cancellation. 
The theoretical  values of Carette {\em et al.}~\cite{Caretal:2013b} for   $A_{3 / 2}$ and $A_{5 / 2}$ of the $2p ^ 4 3p \; ^4\!P ^ o_{3 / 2,5 / 2}$  levels, both strongly affected by relativity, were confirmed experimentally~\cite {Huoetal:2018a}, while there is no experimental values available to compare with for the other two constants $A (2p ^ 4 3s \; ^4\!P_{1 / 2})$ and $A(2p ^ 4 3p \; ^4\!S ^o _{3 / 2} )$.\\
Hyperfine constant values for the $2p^4(^3\!P)3d \;^{2S+1}\!L_J$ levels have been determined recently from  concentration modulation spectroscopy experiments~\cite{Huoetal:2018a}, and it is  worthwhile to investigate  how much relativity affects the  theoretical estimation. The results obtained in the present work far exceeded our expectations, since the relative differences between the nonrelativistic values and those taking relativity into account   reach in some cases  several hundreds percents. As an example, the nonrelativistic correlated values, $A(^4\!F_{3/2})=1333$~MHz, $A(^4\!F_{5/2})=956$~MHz and $A(^4\!F_{7/2})=995$~MHz, 
 are dramatically affected   by the relativistic BP corrections, which decrease them  to  $A(^4\!F_{3/2})=122$~MHz, $A(^4\!F_{5/2})=252$~MHz and $A(^4\!F_{7/2})=263$~MHz, in  good agreement with the experimental values, respectively, $110 \pm 10$, $304 \pm 50$ and $276 \pm 10$~MHz. \\
 
In this work,  we investigate and explain the origin of  the relativistic effects on the calculated hyperfine constants.
We  used the multiconfigurational Hartree-Fock (MCHF)  method to estimate the hyperfine constants of the $2p^4(^3P)3d \; ^{2S+1}\!L_J$ levels, within the framework of a nonrelativistic approach for the optimization of the zero-order wave functions. A simultaneous optimization scheme was applied in the variational nonrelativistic procedure to get a common orbital basis for describing a   set of terms that mix in the Breit–Pauli approximation.
The  relativistic effects are assessed  through Breit-Pauli calculations (MCHF+BP). The latter  are  cross-checked by  relativistic configuration interaction (RCI)  calculations performed in the Pauli approximation.
We also performed fully relativistic MCDHF/RCI calculations based on similar correlation models. These four methods, used for obtaining the relevant electronic wave functions, and  the basic theory of hyperfine interaction, are briefly described in Section~\ref{sec:theo}. 
The simultaneous optimization strategy, used to get a common set of orbitals for the   $2p^4(^3P)3d~^{2S+1}\!L_J$ targeted levels, is described in Section~\ref{par:OS}. The hyperfine constants calculated using the nonrelativistic and relativistic models  are  reported in Section~\ref{sec:cal} for different correlation models and orbital active sets. 
The expressions of the matrix elements of the hyperfine  operators in the configuration state function (CSF) space limited to the $[1/2-9/2]$ range of $J$-values arising from  $2p^4 (^3\!P) 3d \; ^{4,2}(F,D,P)$ terms are fully detailed in Section~\ref{sec:ophfs}.
The theoretical results are analyzed through a detailed comparison with observation in Section~\ref{sec:res}. The main conclusions are resumed in
Section~\ref{sec:concl}.

\section{Theory}\label{sec:theo}

\subsection{Variational methods}
In order to investigate the effects of electronic correlation and relativity on the magnetic dipole hyperfine constant,
we used the multiconfiguration Hartree-Fock (MCHF) approach with Breit-Pauli (BP) corrections and the relativistic configuration interaction method (RCI) in the framework of the Pauli approximation (RCI-P). We also used the multiconfiguration Dirac-Hartree-Fock (MCDHF) method combined with the RCI approach. \\

In the nonrelativistic MCHF method the wave function $\Psi(\gamma \pi LS )$  is a linear combination of configuration state functions (CSFs) $\Phi(\gamma_i \pi LS )$ having the  same parity $\pi$, $L$ and $S$  quantum numbers
 \begin{equation}\label{mchfwf}
\Psi(\gamma \pi LS )=\sum_i c_i \; \Phi(\gamma_i \pi LS ) \;,
 \end{equation}
where the CSFs  are spin-angular-coupled antisymmetric products   of one-electron spin-orbitals $\phi$:
 \begin{equation}\label{one-electron}
 \phi_{nlm_{l}m_{s}} ({\bf r})
=\frac{1}{r}P_{nl}(r)Y_{m_{l}}(\theta, \phi )\chi_{m_{s}} \; .
 \end{equation}
The radial functions $\{P_{n_i l_i}(r)\}$ and the mixing coefficients $\{c_i\}$ in (\ref{mchfwf}) are determined by solving iteratively the numerical MCHF radial equations coupled to the eigenvalue problem in the CSFs space, until  self-consistency. Since the interactions between several of the terms of the $2p^4 3d$ configuration are strong,
 it is important to determine a common set of orbitals for these terms and those that lie  below in the spectrum and have the same parity. In this procedure, referred as simultaneous optimization strategy,  the energy functional is a linear combination of energy functionals for the different $LS$ terms~\cite{Froetal:2007a}.
 Once the one-electron radial functions optimized for the selected states, the BP Hamiltonian matrix is built and diagonalised in the basis of $LSJ$ configuration states belonging to a given parity~$\pi$. The resulting eigenvectors define  the intermediate coupling wave functions  
 \begin{equation}\label{bpwf}
  \Psi(\gamma \pi J) =\sum_k c_k \; \Phi(\gamma_k \pi L_k S_k J) \;,
 \end{equation}
that explicitly illustrates the possible $LS$ mixing for the selected $J$-value. \\
 
 We  also performed relativistic configuration interaction (RCI) calculations to determine the mixing coefficients $\{c_i\}$ of the atomic wave function which, for a state labeled $\gamma \pi J$, is written as a linear combination of relativistic CSFs 
 $\Phi(  \gamma_i \pi J  ) $
\begin{equation}
\label{rciwf}
\Psi (\gamma   \pi J ) = \sum_{i} c_i \; \Phi(  \gamma_i \pi J  ),
\end{equation} 
where the relativistic CSFs  are spin-angular coupled antisymmetric products of one-electron Dirac spinors
\begin{equation}
\label{eq:Dirac_spinor}
\phi_{n \kappa m}({\bf r})=\frac{1}{r}\left(\begin{array}{c}P_{n\kappa}(r)\chi_{\kappa m                            }(\theta,\phi)\\
\mbox{i}Q_{n\kappa}(r)\chi_{-\kappa m}(\theta,\phi)\end{array}\right).
\end{equation}
\\

In the RCI-P method, based on the Pauli limit  of the Dirac equation~\cite{ArmFen:74a}, the radial function of the small component, $Q_{n \kappa}(r) $,  is estimated from the radial function of the large one, $P_{n \kappa}(r)$, as:
\begin{equation}
Q_{n\kappa}(r)  \simeq  \frac{\alpha}{2} \left( \frac{d}{dr} + \frac{\kappa}{r} \right) P_{n \kappa}(r) 
\;,
\end{equation}  
where, in our case,  the large component radial function $P_{n \kappa}(r) $ is the nonrelativistic MCHF radial function $P_{n l}(r) $.\\
In the MCDHF-RCI method, the small and large radial functions of the one-electron Dirac spinors \eqref{eq:Dirac_spinor} are obtained using the fully relativistic MCDHF version of the multiconfiguration method~\cite{Froetal:2016a} to optimise the relativistic one-electron orbital basis, together with the mixing coefficients.

\subsection{Magnetic dipole hyperfine interaction}

The magnetic dipole hyperfine interaction  Hamiltonian  is given by
 \begin{equation}
 H_{\rm{hfs}} =\bm {T}^{(1)} \cdot \bm {M}^{(1)} \; ,
 \end{equation}
 where $\bm{ T}^{(1)}$ is the dipolar magnetic operator tensor which, in the nonrelativistic framework, is the sum of three terms~\cite{LinRos:75a,Hib:75b,Jonetal:93a}
\begin{eqnarray}\label{T1}
 \bm{ T} ^{(1)} = \frac{\alpha^2}{2} \sum_{i=1}^{N}
   \big\{ 2 \bm{ l}^{(1)} (i) r_i^{-3}
  -g_s \sqrt{10} [ \bm{ C}^{(2)}(i) \times \bm{ s}^{(1)}(i) ] ^{(1)} r_i^{-3} 
  + g_s \frac{8}{3} \pi \delta( \bm{ r}_i ) \bm{ s}^{(1)}(i) \big\}
\label{T_1}
\end{eqnarray}
corresponding, respectively, to the orbital, spin-dipole and contact contributions, 
 that we will denote $ \bm{ T} ^{(1)}_{orb}$, $ \bm{ T} ^{(1)}_{sd}$ and $ \bm{ T} ^{(1)}_{con}$, i.e. 
 \begin{equation}
\label{T_split}
\bm{ T} ^{(1)} = 
\bm{ T} ^{(1)}_{orb} +  \bm{ T} ^{(1)}_{sd} + \bm{ T} ^{(1)}_{con} \; .
\end{equation}

The energy corrections of the fine structure levels are generally expressed in term of the magnetic dipole hyperfine constant $A_J$ that is  proportional to the reduced matrix element of $\bm{ T}^{(1)}$
\begin{equation}
\label{A_J_RME}
A_J=\frac{\mu_I}{I}\frac{1}{\sqrt{J(J+1)(2J+1)}}\langle\gamma J \Vert {\bm T}^{(1)} \Vert \gamma J\rangle \; .
\end{equation}
As suggested by Eq.~\eqref{T_split} , 
$A_J$ can be written as 
\begin{equation}
\label{A_J_split}
A_J = A_{J}^{orb} + A^{sd}_{J} + A^{c}_{J} \; .
\end{equation}
where the orbital ($ A_{J}^{orb}$), spin-dipolar ($A^{sd}_{J} $) and contact ($ A^{c}_{J}$) hyperfine constants can be evaluated using \eqref{mchfwf} when omitting relativistic corrections,  and  with \eqref{bpwf} if taking into account relativistic effects through the Breit-Pauli approximation.\\
In the fully relativistic framework of the MCDHF or RCI approaches, the magnetic  electronic tensor operator is (in atomic units) given by~\cite{LinRos:75a,Jonetal:96c}
\begin{equation}
\bm {T}^{(1)} = -  \mbox{i} \; \alpha \sum_{i=1}^{N}\left(\bm \alpha_i \cdot \; \bm l_{i}\bm C^{(1)}(i)\right)r_{i}^{-2},
\end{equation}
and the $ A_{J}$ hyperfine constant (\ref{A_J_RME}) is evaluated  using (\ref{rciwf}).

\section{Simultaneous optimization strategy}  \label{par:OS}

According to the NIST Atomic Spectra Database~\cite{NIST_ASD}, the 17 levels of even parity of interest
\[2p^4 (^3\!P)3d~\; ^4\!D_{7/2,5/2,3/2,1/2}, \; ^2\!D_{5/2,3/2}, \; ^4\!F_{9/2,7/2,5/2,3/2}, \; ^2\!F_{7/2,5/2}, \; ^4\!P_{5/2,3/2,1/2}, \; ^2\!P_{3/2,1/2} \; , \]
arising from the 6 terms $2p^4 (^3\!P)3d~LS$, 
all lie in the narrow spectral window of $[128~064.10 - 128~712.30]$~cm$^{-1}$, above the levels  arising from the 5 terms 
\[ 2p^4 (^3\!P)3s \; ^4\!P, \;  2p^4 (^3\!P)3s \; ^2\!P, \; 2p^4 (^1\!D)3s \; ^2\!D, 
\; 2p^4 (^3\!P)4s \; ^4\!P, \;  2p^4 (^3\!P)4s \; ^2\!P,
\]
of the same parity.
To satisfy the Hylleraas-Undheim-Mac Donald (HUM) theorem~\cite{HylUnd:30a,McD:33a} in the variational procedure, the interaction Hamiltonian matrix should include all low-lying levels of the same $LS$-symmetry in the MCHF procedure. Moreover, because of the orbital orthogonality constraints of the ATSP2K package~\cite{Froetal:2007a}, a single radial orbital basis  has to be obtained for the subsequent BP calculations that mix the levels of the same parity and $J$-value. We therefore adopted a simultaneous optimization scheme~\cite{FroHe:99a,TacFro:99a} for the MCHF calculations, optimizing simultaneously the 6+5=11 terms of even parity. The resulting orbital basis is then used to determine the $J$-dependent energy levels in the framework of the Breit-Pauli approximation.
In the above scheme, the uncorrelated Hartree-Fock (HF) calculation  is done based on the 11 $LS$ terms arising from the $\{2p^4 3d,~2p^4 3s, ~ 2p^4 4s\}$ configuration, and results in a common orthonormal set 
of ``spectroscopic'' orbitals, $(1s,2s,2p,3s,3d,4s)$.  \\

Electron correlation is included by taking the $\{2p^4 3d,~2p^4 3s, ~ 2p^4 4s\}$ configurations as the  multireference (MR),  from which single (S) and double (D) excitations are done to increasing orbital active sets  to build the SD-MR-MCHF expansions. For each orbital active set (AS), all orbitals, spectroscopic and correlation,  are optimized in the MCHF procedure. These calculations are denoted as SD-MR-MCHF[AS], although the latter acronym will be shortened at some places as MR-MCHF[AS], or as (SD)-MR-MCHF[AS] as a discrete reminder, since SD excitations from the MR  are  considered in all the present calculations.
The terminology adapted for the active sets  is detailed in reference~\cite{Aouetal:2018a}. We only recall that the orbital active set (AS) is noted $[n]$ when no angular limitation applies and $[nl]$ when angular orbital limitation $l_{\rm{max}} = l$ is introduced.   \\
The relativistic BP wave function expansions are built using the same SD-MR process, but considering CSFs of all $LS$ symmetries that can be built from the AS and that can mix to each other for a given $J$-value. The corresponding notation, SD-MR-BP[AS], will be used in the following. \\

Table~\ref{tab:E} reports the excitation energies of the  $2p^4 (^3\!P)3d \; ^{2S+1}\!L_J$ levels classified according to the NIST database. As already observed above, the levels lie close to each other. The largest difference between levels having the same $J$-value does not exceed 385~cm$^{-1}$ and is found for the energy separation of  $^2\!D_{5/2}$ and $^4\!F_{5/2}$.  The smallest energy gap, of the order of 90~cm$^{-1}$, is observed between $^2\!F_{5/2}$ and $^4\!P_{5/2}$. \\

 In the same table, the theoretical fine structure  values, $\Delta E_{\text{SD-MR-BP}}$, obtained with the largest  [9f] AS,  are compared with the NIST values. 
For each level, the major contributions to the corresponding  Breit-Pauli wave function are also given. All these contributions correspond to CSFs belonging to the $ 2p ^4 (^3\!P) 3d$ configuration, which form the space that we will indicate in the following as the  $\{ 2p^4 (^3\!P) 3d \; L_iS_i\}$ space. We notice that for all 17 levels, $\sum_i c_i^2\approx 0.97$, illustrating the fact that the CSFs produced by the S and D excitations from the MR only count for around 3\% of the wave functions. The large values of the mixing coefficients clearly demonstrate strong interactions within the $\{ 2p^4 (^3\!P) 3d \; L_iS_i\}$ space.
For example, the contribution of the $ 2p ^ 4 (^ 3\!P) 3d \; ^4\!F_{3/2}, \; ^2\!D_{3/2}, \; ^4\!D_{3/2}, \;^4\!P_{3/2}   $ states  in the composition  of $^2\!P_{3/2}$ level reaches  $\sum c_i^2 - c_1^2 =50.4\% $.
  It is interesting to cite the case of the $^2\!F_{5/2}$ level, which loses its dominant character to the detriment of the $^4\!P_{5/2}$ state with which it strongly interacts. A similar situation has been reported in the case of the $3p^5~4p$ configuration of the argon atom between the $^1\!D_2$ and $^3\!P_2$ states  on the one hand,  and the $^3\!D_1$ and $^1\!P_1$ states on the other~\cite{Dasetal:99a,IriFro:2004a}.
 \\
   A similar simultaneous optimization scheme  was used for the MCDHF calculations, called Extended Optimal Level (EOL) \cite{Graetal:80a},  in which the energy functional is built as the weighted sum of a set of targeted atomic states. With these MCDHF orbital sets, we performed RCI calculations that we note MR-MCDHF-RCI[AS].
  \squeezetable
 \begin{table}
\caption{Excitation energies  according to the NIST Atomic Spectra Database~\cite{NIST_ASD},  fine structures  $\Delta E_{\text{NIST}}$ and $\Delta E_{\text{SD-MR-BP}}$ in cm$^{-1}$  for $2p^4 (^3\!P)3d \; ^{2S+1}\!L_J$ levels, and mixing coefficients of the corresponding SD-MR-BP[9f] eigenvectors.}
\begin{ruledtabular}
\begin{tabular}{ccccrcrccccclcr}\\ 
 Term                     & $ J $  && Level (cm$^{-1}$) & $\Delta E_{\text{NIST}}$ & $\Delta E_{\text{SD-MR-BP}}$ &\multicolumn{9}{c}{Mixing coefficients} \\ [0.1cm]
\hline  \\[-0.2cm]
$ ^{4}$D               &  7/2   && 128 064.10              &   0       & 0     & $0.905~ ^{4}D_{7/2}$   &$+$&$ 0.340~ ^{4}F_{7/2}$&$ +$& $0.193 ~ ^{2}F_{7/2}$&& && \\
                            &  5/2   && 128 087.83               &  23.7  & 22.6 & $0.847 ~^{4}D_{5/2}$  &$ +$&$ 0.326 ~^{4}F_{5/2}$&$ +$& $ 0.248 ~^{4}P_{5/2} $&$ -$& $0.283 ~^{2}D_{5/2}$&$ -$& $ 0.083 ~^{2}F_{5/2}$ \\
                            &  3/2   && 128 122.72               &  58.6 & 56.4  & $0.784~ ^{4}D_{3/2}$  &$ -$&$ 0.399~ ^{2}D_{3/2} $ &$ +$& $ 0.282~ ^{4}P_{3/2} $&$- $& $ 0.259 ~^{2}P_{3/2} $&$ +$& $ 0.226 ~^{4}F_{3/2}$ \\                               
                            &  1/2   && 128 184.99               & 120.9 & 118.6 & $-0.831~ ^{4}D_{1/2}$ &$ +$&$ 0.407~ ^{2}P_{1/2} $ &$ -$& $ 0.338~ ^{4}P_{1/2}$&& && \\
\\
$ ^{2}$D               &  5/2   && 128 140.48 &   0  &0    & $0.827~^{2}D_{5/2}$&$  +$&$ 0.342~ ^{2}F_{5/2}$&$ +$& $ 0.283~ ^{4}P_{5/2} $&$ +$&$0.273~^{4}D_{5/2} $&$ -$& $  0.121 ^{4}F_{5/2}$ \\
                          &  3/2   && 128 219.83 & 79.4 & 78.1  & $-0.717 ~^{2}D_{3/2}$&$  -$&$ 0.345~ ^{2}P_{3/2} $&$ -$& $ 0.464~ ^{4}P_{3/2}$&$ -$& $ 0.338 ~^{4}D_{3/2}  $&$ +$& $  0.090~ ^{4}F_{3/2}$  \\
\\  
 $^{4}$F              &   9/2  &&  128 219.13 &   0  &0  & $  ^{4}F_{9/2}$&&&&&&&& \\
                            &   7/2  &&  128 514.75 & 295.6  & 299.2  &  $0.782~ ^{4}F_{7/2}$&$ +$&$ 0.455~ ^{2}F_{7/2}$&$ -$& $0.391~ ^{4}D_{7/2}$ &&&&  \\
                             &   5/2  &&  128 525.35 & 306.2 & 309.3  & $-0.679~ ^{4}F_{5/2}$&$ +$&$ 0.567~ ^{2}F_{5/2} $&$ -$& $0.387~ ^{2}D_{5/2} $&$ +$&$0.198~ ^{4}D_{5/2} $&$ -$& $0.034~ ^{4}P_{5/2}$ \\
                            &   3/2  &&  128 611.92 & 392.8 & 393.6  & $ 0.821~ ^{4}F_{3/2}$&$ +$&$ 0.514~ ^{2}P_{3/2} $&$ -$& $0.183~ ^{4}P_{3/2} $&$ -$& $ 0.020~ ^{2}D_{3/2} $&$ -$& $ 0.012~ ^{4}D_{3/2}$ \\   
\\   
 $^{2}$F                &   7/2  &&  128 220.36 & 0  & 0&$ 0.853~ ^{2}F_{7/2}$&$ -$&$ 0.495~ ^{4}F_{7/2} $&$ +$& $0.004~ ^{4}D_{7/2}$&&&&  \\
                             &   5/2  &&  128 697.89 &  477.5 & 478.9  & $ -0.442~ ^{2}F_{5/2}$&$ -$& $ 0.671~ ^{4}P_{5/2}$&$ -$& $ 0.362~ ^{4}F_{5/2} $&$ +$& $ 0.372~ ^{4}D_{5/2} $&$ +$& $0.237~ ^{2}D_{5/2}$ \\
\\    
 $^{4}$P                &   1/2  && 128 338.72    & 0   & 0  &$ 0.815~ ^{4}P_{1/2}$&$ +$&$ 0.551~ ^{2}P_{1/2} $&$ -$& $ 0.061~ ^{4}D_{1/2}$&&&& \\
                             &   3/2  &&  128 523.28  & 184.6 & 189.7 &$ 0.762~ ^{4}P_{3/2}$&$ - $&$0.410~ ^{4}D_{3/2} $&$ -$& $0.412~ ^{2}D_{3/2} $&$ +$&$ 0.232~ ^{2}P_{3/2} $&$ +$& $0.008~ ^{4}F_{3/2}$ \\                               
                            &   5/2  &&  128 606.09 &  267.4 & 271.4 & $ -0.614~ ^{4}P_{5/2}$&$ + $&$0.575~ ^{2}F_{5/2} $&$ +$& $0.509~ ^{4}F_{5/2} $&$ +$& $ 0.051~ ^{4}D_{5/2} $&$ +$& $0.030~ ^{2}D_{5/2}$ \\
 \\    
   $^{2}$P              & 1/2  && 128 712.30  &  0     & 0 & $-0.708~ ^{2}P_{1/2}$&$ - $&$0.525~ ^{4}D_{1/2} $&$ +$& $0.440~ ^{4}P_{1/2}$&&&&  \\ 
                           &  3/2  &&  128 520.22  &  192.1 & 192.0  & $ -0.684~ ^{2}P_{3/2}$&$ +$&$ 0.489~ ^{4}F_{3/2}$&$ +$& $0.358~ ^{2}D_{3/2} $&$ -$&$0.274~ ^{4}D_{3/2} $&$ +$& $0.249~ ^{4}P_{3/2}$ \\ \end{tabular}
\end{ruledtabular}
\label{tab:E}
\end{table}

\section{Hyperfine constants calculations}\label{sec:cal}
 $^{19}$F has a nuclear spin  I=1/2 and a nuclear magnetic moment $\mu_I = 2.628868~\mu_{\rm N}$~\cite{Sto:2005a}.
 The magnetic dipole hyperfine constants $A_J$ for all the 17 $2p^4(^3\!P)3d \; ^{2S+1}\!L_J$ levels, calculated using the single- and double-multireference (SD-MR) expansions with the MCHF, BP,  RCI-P and MCDHF-RCI methods, are reported in Tables~\ref{tab:hfs1} and \ref{tab:hfs2}. For the SD-MR-MCHF and SD-MR-BP approaches, the $A_J$ constant value is monitored along the sequence of increasing ASs, from   [4] up to [$9f$], to probe  the correlation effects on the hyperfine structures.
One observes that the hyperfine constant  
values quickly converge with the size of the active space.
Moreover, the $l_{\text{max}}=3$ limitation that has been adopted for building the AS, brings an estimated uncertainty contribution of less than~1\% for the hyperfine constants, deduced by comparing similar calculations performed with [$ng$] active set. 
In other words, the hyperfine constant  
values quickly converge 
not only with the size of the active space, but also with the angular momentum value considered for building the correlation orbital 
active space, a fact that has been observed in many studies, including investigations of the electric field gradient at the nucleus~\cite{SunOls:92b,SunOls:93a}.
From Tables II and III we see that electron correlation effects are small. To highlight this fact, we report in  Table~\ref{tab:rd} the relative difference between  the HF and (SD)-MR-MCHF[9f] hyperfine constants values. This quantity remains smaller than 5\% for nine hyperfine constants and is between $6-14.5\%$  for the others.
Although the description of electron correlation does not seem to be crucial, Tables~\ref{tab:hfs1} 
and~\ref{tab:hfs2} illustrate the large disagreement between the (SD)-MR-MCHF[9f]  theoretical hyperfine constants and the available experimental  values~\cite{Huoetal:2018a},  except for the constant  $A(^2\!D_{3/2})$. 
It becomes clear that the origin of this large theory-observation gap should be found somewhere else than in electron correlation. 
The comparison of the hyperfine constants  between  BP[HF]  and HF (see  Table~\ref{tab:rd}), or between (SD)-MR-BP[9f] and (SD)-MR-MCHF[9f] (not displayed in the Table), indeed indicates huge relativistic effects. The relative differences  reach  values of 1872\%, 898\%, 614\%,  300\%, and 316\%   for, respectively,  $A(^4\!P_{5/2})$, $A(^2\!F_{5/2})$, $A(^4\!F_{3/2})$, $A(^4\!F_{5/2})$, $A(^4\!P_{1/2})$. In the same Table, we  also report the relative differences between (SD)-MR-BP[9f] and BP[HF] hyperfine constants values, which illustrate how much electron excitations beyond the $\{ 2p^4 (^3\!P) 3d \; L_i S_i\}$ space model affects the hyperfine constants.
Except for the four constants $A(^4 \! P_{1/2})$, $A(^2 \! F_{5/2})$, $A(^4 \! P_{5/2})$  and $A(^4 \! F_{3/2})$  for which the corresponding ratio values are large 
(58\%, 20\%, 19\% and 74\% , respectively),
 we observe that  the relativistic effects are efficiently captured through the BP calculations limited to the [HF] active space. For almost all levels considered, the BP[HF] and (SD)-MR-BP[9f]   hyperfine constants  are in good agreement with observation~\cite{Huoetal:2018a}.\\

The  MR-RCI-P[9f] results   are given in  Tables~\ref{tab:hfs1} and \ref{tab:hfs2}.  Since the RCI-P method radially differ from the BP approach, it is interesting to compare the MR-RCI-P[9f] and  MR-BP[9f] hyperfine constant values. We can observe that the two sets of results, obtained using the ATSP2K and GRAPS2018 independent packages, are in excellent agreement with each other. 
In the same tables, we also report the MR-MCDHF-RCI[9f] results. The global agreement of the latter with the MR-BP[9f] results for the 17 hyperfine constants is 4.7\%. The largest differences occur for  $A (^2\!P_{3/2})$, $A (^2\!F_{5/2})$, $A (^4\!F_{3/2})$ and $A (^4\!P_{5/2})$ with relative deviations of 6.8\%, 9.4\%, 19.7\% and 33.3\%, respectively.
 
However, the values obtained in the two approaches  lie within the uncertainty interval of the experimental values for the first constant $A (^2\!P_{3/2})$. This is almost the case for $A (^2\!F_{5/2})$ and $A (^4\!F_{3/2})$,  while the case of the very small $A(^4\!P_{5/2})$ value is more problematic, as it will be further discussed below. 
The global agreement of the averaged MR-BP[9f]/MR-MCDHF-RCI[9f] hyperfine constant values with the 15 available measured hyperfine constants is around  20\%. The largest discrepancies are found for  $A (^4\!P_{1/2})$ and $A (^4\!P_{5/2})$. Excluding the last two from this sample, the global theory-observation agreement drops to 3.5\%. \\

Large differences between the nonrelativistic and Breit-Pauli results are most likely due to the strong relativistic interaction between the terms. In order to verify this conjecture, we   analyse in full details the matrix elements of the different operators of the hyperfine interaction   (see Eqs.~\eqref{T_split} and \eqref{A_J_split}) in the $\{ 2p^4 (^3\!P) 3d \; L_iS_i\}$ space model. It should be noted that the contact hyperfine interaction is zero within this configuration space in which we keep the $1s$ and $2s$ shells closed in the CSF lists. That occupation restriction allows  to limit this detailed analysis to the orbital and spin-dipole contributions to the $A_J$ constants, as done in the next section. However,   the complete hyperfine interaction Hamiltonian is used, including the contact contribution, in the more elaborate calculations based on larger configuration spaces and orbital active sets.

 \begin{table}[H]
\caption{\label{tab:hfs1} Hyperfine structure constants (in MHz) of $2p^4(^3\!P)3d \; ^2D\!$, $^4\!D$ and $^2\!P$  calculated with HF and (SD)-MR-MCHF by using the simultaneous optimization strategy,  BP[HF], (SD)-MR-BP and (SD)-MR-RCI-P methods. These values are compared with fully relativistic results calculated with the (SD)-MCDHF-RCI method, and with observation.} 
\begin{ruledtabular}
\begin{tabular}{ccccccccccc}
&     \multicolumn{2}{c}{$^{2}\!D$} & &    \multicolumn{4}{c}{$^{4}\!D$}     & &\multicolumn{2}{c}{$^{2}\!P$}   \\ [-0.1cm]
 & $A_{3/2}$  & $A_{5/2}$ &&  $A_{1/2}$ &  $A_{3/2}$ & $A_{5/2}$ & $A_{7/2}$ &&    $A_{1/2}$ &    $A_{3/2}$ \\ [+0.1cm]
 \cline{2-3} \cline{5-8} \cline{10-11}\\ [-0.2cm] 
      & \multicolumn{10}{c}{HF}    \\  [-0.2cm] 
      \\
&  1734 &   373  &&  3554  & 1422  &  778 &  169 &&  $-$3346 &  $-$1435 \\ 
 
      & \multicolumn{10}{c}{MR-MCHF}    \\  [-0.2cm] 
      \\
$[4]$ &  1618 &   406  &&  3330  & 1443  &  832 &  228 &&  $-$3345 &  $-$1249 \\ 
$[5f]$ &  1437 &   605  &&  2876  & 1563  & 1027 &  441 &&  $-$3708 &  $-$ 927 \\ 
$[6f]$ &  1744 &   310  &&  3643  & 1434  &  762 &  125 &&  $-$3232 &  $-$1411  \\ 
$[7f]$ &  1678 &   368  &&  3479  & 1454  &  812 &  188 &&  $-$3319 &  $-$1314  \\ 
$[8f]$ &  1674 &   369  &&  3471  & 1451  &  811 &  189 &&  $-$3317 &  $-$1314  \\ 
$[9f]$ &  1675  &  370  &&  3472  & 1453  &  813 &  190 &&  $-$3320 &  $-$1312  \\ \\
      & \multicolumn{10}{c}{BP[HF]}    \\    [-0.2cm] 
      \\
   
&   1574 &  1066  &&  4860  & 2304  & 1474 &  865  &&  $-$2354 &  $-$565 \\ 
      & \multicolumn{10}{c}{MR-BP}    \\    [-0.2cm] 
      \\
$[4]$ &   1579 &  1076  &&  4614  & 2291  & 1494 &  886  &&  $-$2271 &  $-$491 \\
$[5f]$ & 1484   &  1209  &&  4465  & 2523  & 1738 & 1081  &&  $-2134$ &  $-292$  \\ 
$[6f]$ & 1680   &  1033  &&  4733  & 2210  & 1402 &  805  &&  $-2348$ &  $-558$  \\ 
$[7f]$ & 1649   & 1067   && 4658   & 2263  & 1462 & 854   &&  $-$2317 &  $-$506  \\ 
$[8f]$ & 1652   & 1066   && 4647   & 2260  & 1460 & 852  &&  $-$2325 &  $-$503  \\ 
$[9f]$ &  1654  &  1067  && 4646   & 2262  & 1461 & 852  &&  $-$2327 &  $-$496  \\ 
\\
      & \multicolumn{10}{c}{MR-RCI-P}    \\    [-0.2cm] 
      \\
$[9f]$ & 1652   &  1065  &&  4640  & 2258  & 1458 & 850  &&  $-$2326   & $-$497  \\      
\\
      & \multicolumn{10}{c}{MR-MCDHF-RCI}    \\    [-0.2cm] 
      \\
$[9f]$ & 1649  &  1066  &&  4608  & 2257  & 1463 & 855  &&  $-$2312   & $-$463  \\ 
\\
Exp~\cite{Huoetal:2018a}            &   $1582 \pm 50$ &  $1046 \pm 50$ && $4541 \pm 50$ &   $2290 \pm 50$ &  $1481 \pm 20$ &   $793 \pm 20$ &&   $-2378 \pm 80$ & $-498 \pm 80$ \\
\end{tabular}
\end{ruledtabular}
\end{table}
\begin{table}[H]
\caption{\label{tab:hfs2} Hyperfine structure constants in (MHz) of $2p^4(^3\!P)3d \; ^4\!P$, $^2\!F$ and $^4\!F$  calculated 
with HF and (SD)-MR-MCHF by using the simultaneous optimization strategy, BP[HF], (SD)-MR-BP and (SD)-MR-RCI-P methods. These values are compared with fully relativistic results calculated with the (SD)-MCDHF-RCI method, and with observation. }
\begin{ruledtabular}
\begin{tabular}{cccccccccccc}
 &      \multicolumn{3}{c}{$^{4}\!P$ }  && \multicolumn{2}{c}{$^{2}\!F$}   &&  \multicolumn{4}{c}{$^{4}\!F$}  \\[-0.1cm]
   &  $A_{1/2}$ & $A_{3/2}$ & $A_{5/2}$ && $A_{5/2}$ & $A_{7/2}$&& $A_{3/2}$ &$A_{5/2}$ & $A_{7/2}$ & $A_{9/2}$  \\[+0.1cm]
 \cline{2-4} \cline{6-7} \cline{9-12}\\[-0.2cm] 
  & \multicolumn{11}{c}{HF}    \\  [-0.2cm] 
  \\
   & 1574  &  $-$ 724 & $-$848 && 1285  & 1437 && 1515 & 1021  & 1015  & 1184 \\
       &   \multicolumn{11}{c}{MR-MCHF}  \\
       \\[-0.2cm] 

$[4]$ & 1823  &  $-$ 565 & $-$699 && 1156  & 1439 && 1295 &  962  & 1010  & 1207 \\
$[5f]$ & 2667  &  $-$ 206 & $-$406 && 1017  & 1594 &&  995 &  978  & 1135  & 1384 \\
$[6f]$ & 1460  &  $-$ 741 & $-$841 && 1218  & 1390 && 1416 &  950  &  961  & 1146 \\
$[7f]$ & 1694  &  $-$ 634 & $-$755 && 1178  & 1427 && 1334 &  956  &  994  & 1190 \\
$[8f]$ & 1693  &  $-$ 634 & $-$755 && 1180  & 1427 && 1337 &  957  &  994  & 1189 \\
$[9f]$ & 1699  &  $-$ 632 & $-$753 && 1178  & 1428 && 1333 &  956  &  995  & 1191 \\
\\
& \multicolumn{11}{c}{BP[HF]}    \\    [-0.2cm] 
\\
 & $-$730   &  $-$1029  & $-$43 && $-$161 &  1457 &&  212 &  255 &  282  & 1176 \\  
 \\
       &   \multicolumn{11}{c}{MR-BP}  \\
       \\[-0.2cm] 

$[4]$ & $-$530   &  $-$ 995  & $-$30 && $-$174 &  1494 &&  149 &  267 &  280  & 1201 \\
$[5f]$ & $-491$   &  $-910$   & $-39$ && $-$ 93 &  1716 &&   48 & 
383 &  353 &  1377 \\
$[6f]$ & $-509$  &  $-1048$   & $-41$ && $-$220 &  1420 &&  146 & 
201 &  230 &  1139 \\
$[7f]$ & $-$482  &  $-$1028 & $-$34 && $-$203 & 1477 &&  130  & 244   & 258  & 1184  \\
$[8f]$ & $-$469  &  $-$1033 & $-$32 && $-$205 & 1477 &&  129  &  250 &  262 & 1184 \\
$[9f]$ & $-$462  &  $-$1033 & $-$36 && $-$202 & 1480 &&  122  & 252  & 263  &  1185\\
\\
       & \multicolumn{11}{c}{MR-RCI-P}    \\    [-0.2cm] 
      \\
$[9f]$ & $-$461 &   $-$1033 & $-$34 && $-$202 & 1478 &&  124  & 252 & 263 &  1183\\    
\\
 & \multicolumn{11}{c}{MR-MCDHF-RCI}    \\    [-0.2cm] 
      \\
$[9f]$ & $-$445 &   $-$1026 & $-$48 && $-$183 & 1483 &&  98  & 259 & 266 &  1188\\ 
\\
 Exp~\cite{Huoetal:2018a}          &   $-226 \pm 50$& $-1035 \pm 50$ & $-17 \pm 10$ &&   $-190 \pm 10$     &     &&    $110 \pm 10$ &  $304 \pm 50$  &  $276 \pm 10$ &  \\      
\end{tabular}
\end{ruledtabular}
\end{table}

\begin{table}[H]
\caption{Relative differences in percent between (MR-MCHF[9f], HF), (BP[HF], HF), and (MR-BP[9f], BP[HF]) hyperfine constants}
\begin{ruledtabular}
\begin{tabular}{cccccccccccccccccccccc}\\ [-0.2cm]
 \multicolumn{2}{c}{$^{2}\!D$ } &&  \multicolumn{4}{c}{$^{4}\!D$ } &&  \multicolumn{2}{c}{$^{2}\!P$ } &&  \multicolumn{3}{c}{$^{4}\!P$ } & & \multicolumn{2}{c}{$^{2}\!F$ } &&  \multicolumn{4}{c}{$^{4}\!F$ } \\

$A_{3/2}$  & $A_{5/2}$ &&  $A_{1/2}$ &  $A_{3/2}$ & $A_{5/2}$ & $A_{7/2}$ &&    $A_{1/2}$  &    $A_{3/2}$ &&  $A_{1/2}$ & $A_{3/2}$ & $A_{5/2}$ && $A_{5/2}$ & $A_{7/2}$&& $A_{3/2}$ &$A_{5/2}$ & $A_{7/2}$ & $A_{9/2}$  \\[+0.2cm]
 
\hline
\\[-0.1cm]
\multicolumn{22}{c}{$\frac{\vert \text{MR-MCHF[9f]} - \text{HF} \vert}{\vert \text{MR-MCHF[9f]}\vert } $}\\
\\[-0.2cm]
3.5 &  0.8 && 2.4 & 2.1& 4.3 & 11 && 0.8 & 9.3 && 7.3 & 14.5 & 12.6 && 9.0 & 0.6 && 13.6 & 6.7 & 2.0 & 0.5 \\
\\[-0.2cm]
\hline
\\[-0.1cm]
  \multicolumn{22}{c}{$\frac{\vert \text{BP[HF]} - \text{HF} \vert}{\vert \text{BP[HF]}\vert } $}\\
\\[-0.2cm]
10.1 &  65.0 && 26.8 & 38.2& 47.2 & 80.4 && 42.1 & 153.9 && 315.6 & 29.6 & 1872 && 898.1 & 1.3 && 614 & 300.3 & 259.9 & 0.6 \\
\\[-0.2cm]
\hline
\\[-0.1cm]
  \multicolumn{22}{c}{$\frac{ \vert \text{MR-BP}[9f]-  \text{BP[HF]} \vert}{ \vert \text{MR-BP}[9f] \vert }   $}\\
\\[-0.2cm]
4.8 & 0.0  && 4.6 & 1.8 & 0.9 & 1.5 && 1.2 & 13.9 && 58.0 & 0.4 & 19.4 && 20.2 & 1.5 && 73.7 & 1.2 & 7.2 & 0.7 \\
\end{tabular}
\end{ruledtabular}
\label{tab:rd}
\end{table}

\section{M1 hyperfine interaction in the $\{ 2p^4 (^3\!P) 3d \; L_iS_i\}$ space}

\subsection{Matrix elements}\label{sec:ophfs}

In the present section, we limit the CSFs to  the  $\{ 2p^4 (^3\!P) 3d \; L_iS_i\}$ space. The atomic wave function describing the $\gamma LSJ$  states, where $\gamma = 2p^4(^3\!P)3d$,  are therefore written according to Eq.~(\ref{bpwf}) as follows:
\begin{equation}
\Psi\left(\gamma~^{2S+1}\!L_J\right)=\sum_i c_i \; \phi\left(\gamma~L_i S_i J\right) \; ,
\end{equation}
where  $L_i S_i$ represents any of the six   terms listed in Table~\ref{tab:E} corresponding to the same  $J$-value.
In this approximation, that keeps the $1s$ and $2s$ shells closed, there is no contact contribution and the hyperfine constant  $A(^{2S+1}\!L_{J})$ of each level
$2p^4 (^3P) 3d \; LSJ$ 
 is only made of 
 the orbital and spin-dipole contributions, i.e.
\begin{equation}\label{AJ}
A(^{2S+1}\!L_{J}) 
= A^{orb}(^{2S+1}\!L_{J}) 
+ A^{sd}(^{2S+1}\!L_{J}),
\end{equation}
where
 \begin{equation}\label{AJorb-sd}
 A^{orb}(^{2S+1}\!L_{J}) = \sum_{ij } A_{J}^{orb}(\gamma L_i S_i, \gamma L_j S_j) \;\;\; \text{and} \;\;\;   
 A^{sd}(^{2S+1}\!L_{J}) = \sum_{ij} A_{J}^{sd}(\gamma L_i S_i,\gamma L_j S_j)
\end{equation}
are made of the diagonal ($i=j$) and off-diagonal ($i \ne j$) hyperfine interaction matrix elements coupling the  CSFs in the  basis. 
  $ A_{J}^{orb}(\gamma L_i S_i,\gamma L_j S_j)$  and  $ A_{J}^{sd}(\gamma L_i S_i,\gamma L_j S_j)$  are proportional, respectively, to the reduced matrix elements $\langle\gamma L_i S_iJ \Vert {\bm T}^{(1)}_{orb} \Vert \gamma L_j S_jJ\rangle$ and $\langle\gamma L_i S_iJ \Vert {\bm T}^{(1)}_{sd} \Vert \gamma L_j S_j J\rangle$ and to the relevant eigenvector coefficient products $c_ic_j$. They can be written as 
 \begin{equation}\label{AJorb-sd-ij}
A_{J}^{orb}(\gamma L_i S_i,\gamma L_j S_j) = \frac{1}{2} c_{i}c_{j} G_{\mu}\frac{\mu_{I}}{I}
\text{ME}^{orb} \;\;\;\;   \text{and}     \;\;\;  A_{J}^{sd}(\gamma L_i S_i,\gamma L_j S_j) = \frac{1}{2} c_{i}c_{j} G_{\mu}\frac{\mu_{I}}{I}
\text{ME}^{sd} \; ,
\end{equation}
with~\cite{Arm:71a,Jonetal:93a}
\begin{multline}
\label{eq:ME_orb}
\text{ME}^{orb} =\delta_{S_i S_j} (-1)^{L_{i}+S_{i}+J+L_{j}+1}\; 2 \; \sqrt{\frac{(2L_{i}+1)(2L_{j}+1)(2J+1)}{J(J+1)}}
\ \left\{
\begin{array}{ccc}
 L_i & S_i & J \\
 J & 1& L_j
\end{array}
 \right\}\\ 
\times \left[
 \left\{
\begin{array}{ccc}
 1& 2 & L_i \\
 L_j & 1& 1
\end{array}
 \right\}\sqrt{6} \langle 2p^{4}~^{3}P \|{\bf U}^{(1)}\|2p^{4}~^{3}P \rangle \langle 2p |r^{-3}|2p\rangle +
 \left\{
\begin{array}{ccc}
 2 & 1 & L_i \\
L_j & 1 & 2
\end{array}
 \right\}
 \sqrt{30} \langle 3d~^{2}D \|{\bf U}^{(1)}\|3d~^{2}D \rangle \langle 3d |r^{-3}|3d\rangle
 \right] \; ,
\end{multline}
\\ \noindent and 
\begin{multline}
\label{eq:ME_sd}
\text{ME}^{sd} =
(-1)^{ S_j + L_j+1/2}
\sqrt{\frac{(2S_i+1)(2S_j+1)(2L_i+1)(2L_j+1)(2J+1)}{J(J+1)}}g_s\sqrt{30}
\left\{
\begin{array}{ccc}
 L_i & S_i & J \\
L_j & S_j & J\\
 2   &  1  &  1
\end{array}
 \right\} \\ 
\times \Bigg[-\sqrt{\frac{6}{5}}
 \left\{
\begin{array}{ccc}
 1 & 1/2 & S_i \\
S_j & 1 & 1
\end{array}
 \right\} 
 \left\{
\begin{array}{ccc}
 1 & 2 & L_i \\
L_j & 2 & 1
\end{array}
 \right\} 
 \langle 2p^{4}~^{3}P \|{\bf V}^{(21)}\|2p^{4}~^{3}P \rangle  
\langle 2p |r^{-3}|2p\rangle \\
-\sqrt{\frac{10}{7}}
 \left\{
\begin{array}{ccc}
 1/2 & 1 & S_i \\
S_j & 1 & 1/2
\end{array}
 \right\} 
 \left\{
\begin{array}{ccc}
 2 & 1 & L_i \\
L_j & 2 & 1
\end{array}
 \right\} 
 \langle 3d~^{2}D \|{\bf V}^{(21)}\|3d~^{2}D  \rangle   
 \langle 3d |r^{-3}|3d\rangle
 \Bigg] \; .
 \end{multline}
$G_{\mu} = 95.41068$ is the  numerical factor to be used when expressing $\text{ME}^{orb}$ and $\text{ME}^{sd}$ in atomic units ($a_0^{-3}$), $\mu_I$ in nuclear magnetons ($\mu_{\rm N}$) and $A_J$ in units of frequency (MHz) while $g_s = 2.0023193$ is the electronic $g$ factor corrected for the quantum electrodynamic (QED) effects. ${\bf U}^{(1)}$  is the unit tensor operator acting only in the $L$-space, and ${\bf V}^{(21)}$ is the unit double tensor operator~\cite{Cow:81a}. $\langle nl |r^{-3}|nl \rangle$ are the one-electron radial integrals for the active subshells, $nl=2p$ and $3d$. 
The numerical factors $\sqrt{6}$ and $\sqrt{30}$, appearing in~(\ref{eq:ME_orb}) correspond to the reduced matrix elements of the angular momentum operator $\langle l \Vert {\bf l}^{(1)}\Vert l \rangle$ for $l=1$ and $l=2$, respectively. 
In the same way, the numerical factors $-\sqrt{6/5}$ and $-\sqrt{10/7}$, appearing in~(\ref{eq:ME_sd}) correspond to the reduced matrix elements of the renormalized spherical harmonic $\langle l \Vert {\bf C}^{(2)} \Vert l \rangle$ for $l=1$ and $l=2$ (compare the structure of eqs.~(27) and (31) in ~\cite{Jonetal:93a}).

\subsection{Detailed analysis}\label{sec:res}

The numerical values of the products of the mixing coefficients    $c_i c_j$, the
 electronic matrix elements,
ME$^{orb}$ (\ref{eq:ME_orb}) and  ME$^{sd}$~(\ref{eq:ME_sd}), 
as well as the results of the formulas (\ref{AJ}), (\ref{AJorb-sd}) and (\ref{AJorb-sd-ij}) are reported in Tables~\ref{tab:A_2D32} and  \ref{tab:A_2D52}, for $2p^{4}(^{3}P)3d~^{2}D_{3/2}$ 
 and
 $2p^{4}(^{3}\!P)3d~^{2}\!D_{5/2}$, 
 respectively.
  The mixing coefficients of the corresponding eigenvectors are taken from the MR-BP[9f] calculations (see Table~\ref{tab:E}).  The resulting $ A_{J}^{orb}(\gamma L_i S_i,\gamma L_j S_j)$ and $ A_{J}^{sd}(\gamma L_i S_i,\gamma L_j S_j)$ values are given in the fourth and sixth columns, respectively.
For each ($L_iS_i,L_jS_j$) relevant pair,
the sum of the orbital and spin-dipolar contributions is reported in the very last column.
At the bottom of the table, we give the total values of the orbital and spin-dipolar hyperfine constants, together with their resulting sum  respectively,
from the contribution of the matrix elements in the $\{ 2p^4 (^3\!P) 3d \; L_iS_i\}$ space and from HF and MR-BP[9f] calculations. 
As already indicated previously, the hyperfine contact interaction is strictly zero in the  $\{ 2p^4 (^3\!P) 3d \; L_iS_i\}$ \mbox{space,} but not anymore in the spaces associated with the HF calculations in the simultaneous optimisation scheme for $2p^4 (^3\!P) 3d \; ^2\!P$, $^4\!P$, and $^2\!D$ states, 
that involve the contamination by CSFs with one electron $3s$ or $4s$. The same observation can be done for  the (SD)-MR-BP[9f] calculations for all states $2p^4 3d$ $LS$,
for which the opening of the $1s^2$ and $2s^2$ subshells  switches on the contact contribution through the spin-polarization excitation mechanism \cite{LinMor:82a}.
The latter, however,
remains rather small. Indeed, as one can see in the two Tables~\ref{tab:A_2D32} and  \ref{tab:A_2D52},  the contribution of the contact interaction does not exceed 1\% in the HF  calculations and is of the order of 2\% in the MR-BP[9f] calculations. The experimental values are given in the last line.  \\
  
  The two tables illustrate the large  effects of  terms mixing   on the orbital and spin-dipole constants through the factors $c_i c_j$.  For example for the   state $^2\!D_{3/2}$,  the contributions to the orbital hyperfine constant  of two non-diagonal matrix elements, ($^2\!D$,$^2\!P$) and ($^4\!P$, $^4\!D$), which are respectively equal to 428~MHz and 487~MHz  are of the same order of magnitude as that of the main matrix element  ($^2\!D$,$^2\!D$)  which is 447~MHz. The total contribution  of the mixing states  to the constant $A^{orb}(^2\!D_{3/2}) $  is 646~MHz,  or 59\% of  a total of 1093~MHz,  despite a compensation effect estimated to 354~MHz, due to the mixing with other  $LS$-component. Mixing effects on the  spin-dipolar $A^{sd}(^2\!D_{3/2}) $  constant are reduced by cancellation effects. Their contribution to the total hyperfine constant is of the order of 47\%. The term-mixing effect on the total hyperfine constants  depend on the relative sign of the orbital and spin-dipole contributions resulting from each matrix element. They are often reduced due to opposite signs, inducing strong cancellation. In the case of $A(^2\!D_{3/2})=1618$~MHz, these effects are of the order of 47\%. Finally, the value of $A(^2\!D_{3/2})$ obtained using the $\{ 2p^4 (^3\!P) 3d \; L_iS_i\}$ space represents 98\% of the value resulting from the MR-BP[9f] calculation,  which is based on a  space formed by 1~114~108 CSFs. We then deduce that most of the relativistic effects due to mixing effects are captured by the single $\{ 2p^4 (^3\!P) 3d \; L_iS_i\}$ space. The  results corresponding to the two calculations BP[HF] and MR-BP[9f] are in  good agreement with the experiment.\\
  
   For the level $^{2}\!D_{5/2}$ (Table~\ref{tab:A_2D52}), the HF hyperfine constants values, $A_{5/2}^{orb}$ and $A_{5/2}^{sd}$,    change, respectively,  from 607 to 827~MHz and from $-$236 to 186~MHz, when using  the BP[HF] model, equivalent to a variation of the total $A(^2D_{5/2})$ constant from 371~MHz to 1013~MHz. We notice a particularly important effect on the spin-dipole interaction.  This effect is mainly due to the two matrix elements ($^2\!D$,$^2\!F$),  ($^2\!D$, $^4\!D$) of the spin-dipole operator, which increase the spin-dipole contribution, respectively, by 280~MHz and 119~MHz. 
Note that  among the  $^{2}\!D_{5/2}$ eigenvector $LS$-composition,  the contribution of $^{2}\!F_{5/2}$ to the constant $A_{5/2}$ is  641~MHz, which corresponds to 63\% of the total value. \\

In Tables~\ref{tab:J32}-\ref{tab:J72} we report in details, for all the other considered levels, the contributions of the hyperfine orbital ($orb$), spin-dipolar ($sd$) constants, their sum ($orb + sd$) for each matrix element, as well as the totals $A_J^{orb}$, $A_J^{sd}$, and $A_J$. In the penultimate row we report the  MR-BP[9f] values, that  we compare with observation~\cite{Huoetal:2018a} in the last row, when available. \\

The value of the $c_1$  coefficient in the development of the wave functions from Table~\ref{tab:E} is a good indicator of the importance of the  relativistic effects.
If the coefficients $c_i$ are  deduced from a Breit-Pauli calculation limited to the $\{ 2p^4 (^3\!P) 3d \; L_iS_i\}$ space such that $\sum_i c_i^2 =1$, the weight $c^2_1$ can be written as follows:
 \begin{equation}\label{c1}
c^2_1=\frac{A_{J}^{orb}(LS,LS)}{  A^{orb}(^{2S+1}\!L_{J})\text{[HF]} }
= \frac{A_{J}^{sd}(LS,LS)}{  A^{sd}(^{2S+1}\!L_{J})\text{[HF]} }
= \frac{A_{J}(LS,LS)}{ A(^{2S+1}\!L_{J})\text{[HF]} } \; ,
\end{equation}
where $c^2_1=1$  would correspond to a  Hartree-Fock calculation.  
When  $\vert c_1 \vert$ decreases, the relative difference between  $A_{J}(LS,LS)$ and 
$  A(^{2S+1}\!L_{J})\text{[HF]} $ increases, which reveals large term-mixing effects. 
This can be illustrated in the case of  $^{2}\!F_{5/2}$ for which  $\vert c_1 \vert =0.442$ (see Table~\ref{tab:E}),
with the following values: $A_{5/2}^{orb}(^2\!F,~^2\!F)= 323$~MHz,  $A_{5/2}^{sd}(^2\!F,~^2\!F)=~-77$~MHz,  $A_{5/2}(^2\!F,~^2\!F)= 246$~MHz and 
$  A^{orb}(^2\!F_{5/2})\text{[HF]}=1691$~MHz, 
$  A^{sd} (^2\!F_{5/2})\text{[HF]}=-406$~MHz, 
$  A(^2\!F_{5/2})\text{[HF]}=1285$~MHz (see Table~\ref{tab:J52}). We can observe however that the relations~(\ref{c1}) are not perfectly verified because  the $c_i$ coefficients reported in Table~\ref{tab:E} are taken from the SD-MR-BP[9f] eigenvectors  and therefore do not fully satisfy $\sum_i c_i^2 = 1$. The large difference between the two values of $A_{5/2}(^2\!F,~^2\!F)$ and $ A(^2\!F_{5/2})\text{[HF]} $ indicates a significant contribution from the other matrix elements, as it can be seen in Table~\ref{tab:J52}  (column 10 entitled ``$orb + sd$~''). \\

For all states, the hyperfine constants calculated using MR-BP[9f] or MR-RCI-P[9f] agree very well with observation, except for $A(^4\!P_{1/2})$, and $A(^4\!P_{5/2})$ , as already commented at the end of Section~\ref{sec:cal}. 
For the first case ($A(^4\!P_{1/2})$), Table~\ref{tab:J12} illustrates a huge cancellation between the two diagonal contributions,
$A_{1/2}(^2\!P,~^2\!P)= -993$~MHz and $A_{1/2}(^4\!P,~^4\!P)= 1022$~MHz, leaving much room to the off-diagonal coupling matrix element $A_{1/2}(^4\!P,~^4\!D)=-516$~MHz.
For the second case ($A(^4\!P_{5/2})$), the fact that this hyperfine constant is the smallest one (in absolute value) amongst the 15 experimental values can be easily understood from the very large cancellation between the orbital and spin-dipole contributions, as demonstrated by Table~\ref{tab:J52}.
The use of the $\{ 2p^4 (^3\!P) 3d \; L_iS_i\}$ space combined with the $c_i$ coefficients of the MR-BP[9f] eigenvector made it possible to demonstrate very clearly the effects of the term-mixing  on the hyperfine constants. In some cases, like $^4\!F_{3/2}$ and $^4\!P_{3/2}$ for example, the $\{ 2p^4 (^3\!P) 3d \; L_iS_i\}$ limited space is not large enough to obtain a good agreement with the [9f]-space result, but is sufficient to demonstrate the importance of the mixtures.

\begin{table}[H]
\caption{\label{tab:A_2D32} Values of $ A_{3/2}^{orb}(\gamma L_i S_i,\gamma L_j S_j)$,    $A^{orb}(^{2}\!D_{3/2})$,   $ A_{3/2}^{sd}(\gamma L_i S_i,\gamma L_j S_j)$    $A^{sd}(^{2}\!D_{3/2}) $,  $ A_{3/2}(\gamma L_i S_i,\gamma L_j S_j)$ and $ A(^{2}\!D_{3/2})$ in MHz according to the  formulas~(\ref{AJ}), (\ref{AJorb-sd}), (\ref{AJorb-sd-ij}).}
\begin{ruledtabular}
\begin{tabular}{lrrrrrrr}
$ (L_i S_i,  L_j S_j)$       & $c_i c_j$    &$\text{ME}^{orb}$  & $ A_{3/2}^{orb}(\gamma L_i S_i,\gamma L_j S_j)$ &  $\text{ME}^{sd}$   &$ A_{3/2}^{sd}(\gamma L_i S_i,\gamma L_j S_j)$   & $ A_{3/2}(\gamma L_i S_i,\gamma L_j S_j)$  \\
\hline             
($^{2}D$,$^{2}D$ ) & 0.5145 &    3.4612 &    447 &   3.2288 &      417 &      864 \\    
($^{2}P$,$^{2}P$ ) & 0.1187 & $-$5.7542 & $-$171 &   0.1539 &        5 &  $-$ 166 \\          
($^{4}P$,$^{4}P$ ) & 0.2156 & $-$2.3016 & $-$124 &$-$0.5220 & $-$   28 &  $-$ 152 \\   
($^{4}F$,$^{4}F$ ) & 0.0081 &    9.2212 &     19 &$-$3.3199 & $-$    7 &       12 \\ 
($^{4}D$,$^{4}D$ ) & 0.1143 &    2.3074 &     66 &   3.2298 &       93 &      159 \\
2$\times$($^{2}D$,$^{2}P$ ) & 2$\times$0.2472    &   3.4540 &      428 & $-$1.3844 &  $-$  172 &     256 \\
2$\times$($^{2}D$,$^{4}P$ ) & 2$\times$0.3330    &    00    &      00  &    0.1549 &        26 &      26 \\ 
2$\times$($^{2}D$,$^{4}F$ ) & 2$\times$ $-$0.0645&    00    &      00  & $-$2.3147 &        75 &      75 \\
2$\times$($^{2}D$,$^{4}D$ ) & 2$\times$0.2425    &    00    &      00  &    0.8069 &        98 &      98 \\
2$\times$($^{2}P$,$^{4}P$ ) & 2$\times$0.1600    &    00    &      00  & $-$0.2242 &  $-$   18 & $-$  18 \\
2$\times$($^{2}P$,$^{4}F$ ) & 2$\times$ $-$0.0310&    00    &      00  & $-$1.1580 &        18 &      18 \\
2$\times$($^{2}P$,$^{4}D$ ) & 2$\times$0.1165    &    00    &      00  &    1.0390 &        61 &      61 \\
2$\times$($^{4}P$,$^{4}D$ ) & 2$\times$0.1570    &   6.1800 &     487  &        00 &        00 &     487 \\ 
2$\times$($^{4}F$,$^{4}D$ ) & 2$\times$ $-$0.0304&   3.8538 & $-$  59  &        00 &        00 & $-$  59 \\ 
2$\times$($^{4}P$,$^{4}F$ ) & 2$\times$ $-$0.0417&    00    &      00  &    2.0711 &   $-$  43 & $-$  43 \\ 
\hline  \\[-0.1cm]
   & & &$ A^{orb}(^{2}\!D_{3/2})$ & & $ A^{sd}(^{2}\!D_{3/2})$ & $A(^{2}\!D_{3/2})$   \\\\[-0.2cm]
       & &  &  1093 &  & 525 &  1618        \\\hline \\[-0.2cm]
 HF                          & &  &910  &   & 825 &  1734$\dag$        \\
   \\[-0.2cm]
MR-BP[9f]                         & &  &1118  &   & 563 &  1654$\dag$      \\
 \\[-0.2cm]
Exp \cite {Huoetal:2018a}     & &  &       &  &     & $1582 \pm 50$  \\
\end{tabular}
$\dag$ These totals  differ from $ A^{orb}(^{2}D_{3/2})  +  A^{sd}(^{2}D_{3/2}) $ because they include the contact contribution, which is not strictly zero in the  HF   and MR-BP[9f]  calculations (see text for more details). \\
\end{ruledtabular}
\end{table} 

\begin{table}[H]
\caption{\label{tab:A_2D52} Values of $ A_{5/2}^{orb}(\gamma L_i S_i,\gamma L_j S_j)$,    $A^{orb}(^{2}D_{5/2})$,   $ A_{5/2}^{sd}(\gamma L_i S_i,\gamma L_j S_j)$    $A^{sd}(^{2}D_{5/2}) $,  $ A_{5/2}(\gamma L_i S_i,\gamma L_j S_j)$ and $ A(^{2}D_{5/2})$ in MHz according to the formulas ~(\ref{AJ}), (\ref{AJorb-sd}), (\ref{AJorb-sd-ij}).}
\begin{ruledtabular}
\begin{tabular}{lrrrrrrr}
$ (L_i S_i,  L_j S_j)$        & $c_i c_j$    &$\text{ME}^{orb}$    & $ A_{5/2}^{orb}(\gamma L_i S_i,\gamma L_j S_j)$ &  $\text{ME}^{sd}$   &$ A_{5/2}^{sd}(\gamma L_i S_i,\gamma L_j S_j)$   &$ A_{5/2}(\gamma L_i S_i,\gamma L_j S_j)$ \\[+0.2cm]
\hline \\[-0.2cm]           
($^{2}D$,$^{2}D$ ) & 0.6846 &   2.3074 &      396 & $-$0.9225 & $-$ 158 &      238 \\    
($^{2}F$,$^{2}F$ ) & 0.1170 &    6.5866 &      193 & $-$1.5820 & $-$  46 &      147 \\          
($^{4}P$,$^{4}P$ ) & 0.0804 & $-$3.4526 & $-$   70 &    0.1382 &       3 & $-$   67 \\   
($^{4}F$,$^{4}F$ ) & 0.0145 &    5.5986 &       20 & $-$1.6204 & $-$   6 &       14 \\ 
($^{4}D$,$^{4}D$ ) & 0.0748 &    1.8130 &       34 &    1.2194 &      23 &       57 \\
2$\times$($^{2}D$,$^{2}F$ ) & 2$\times$     0.2831 &    1.2310 &     175 &    1.9724 &     280 &     455 \\
2$\times$($^{2}D$,$^{4}P$ ) & 2$\times$     0.2346 &        00 &      00 & $-$0.3794 &  $-$ 45 & $-$  45 \\ 
2$\times$($^{2}D$,$^{4}F$ ) & 2$\times$ $-$ 0.0998 &        00 &      00 & $-$1.2125 &      61 &      61 \\
2$\times$($^{2}D$,$^{4}D$ ) & 2$\times$     0.2263 &        00 &      00 &    1.0475 &     119 &     119 \\
2$\times$($^{2}F$,$^{4}P$ ) & 2$\times$     0.0970 &        00 &      00 &    0.4052 &      20 &      20 \\
2$\times$($^{2}F$,$^{4}F$ ) & 2$\times$ $-$ 0.0413 &        00 &      00 & $-$0.6195 &      13 &      13 \\
2$\times$($^{2}F$,$^{4}D$ ) & 2$\times$     0.0936 &        00 &      00 &    0.1317 &       6 &       6 \\
2$\times$($^{4}P$,$^{4}D$ ) & 2$\times$     0.0775 &    3.0342 &     118 & $-$1.0124 &  $-$ 39 &      79 \\ 
2$\times$($^{4}F$,$^{4}D$ ) & 2$\times$ $-$ 0.0330 &    2.3542 & $-$  39 &    1.1791 &  $-$ 20 &  $-$ 59 \\ 
2$\times$($^{4}P$,$^{4}F$ ) & 2$\times$ $-$ 0.0342 &        00 &      00 &    1.4495 &  $-$ 25 &  $-$ 25 \\
\hline  \\[-0.1cm]
  & & &  $ A^{orb}(^{2}D_{5/2})$ & & $ A^{sd}(^{2}D_{5/2})$  & $A(^{2}D_{5/2})$   \\\\[-0.2cm]
   & &  &  827 &  & 186 & 1013    \\\hline \\[-0.2cm]
      HF                         & &  &607  &   & $-$236 &  373$\dag$       \\
 \\[-0.2cm]
MR-BP[9f]                         & &  &  843 &  & 197 & 1067$\dag$      \\
 \\[-0.2cm]
Exp \cite{Huoetal:2018a}    & & &      &  &   &  $1046 \pm 50$   \\
\end{tabular}
$\dag$ These totals  differ from $ A^{orb}(^{2}D_{5/2})  +  A^{sd}(^{2}D_{5/2}) $ because they include the contact contribution which is not strictly zero in the  HF   and MR-BP[9f]  calculations (see text for more details). \\
\end{ruledtabular}
\end{table}

\begin{table}[H]
\caption{\label{tab:J12} Values of  $ A^{orb}_{1/2}(\gamma L_i S_i,\gamma L_j S_j)=orb$,  $ A^{sd}_{1/2}(\gamma L_i S_i,\gamma L_j S_j)=sd$, $ A_{1/2}(\gamma L_i S_i,\gamma L_j S_j)=orb + sd$ for  $^{4}\!D$ ,   $^{4}\!P$   and $^{2}\!P$ states. At the bottom of the table we give the total values $A^{orb}_{1/2}$, $A^{sd}_{1/2}$ and  $A_{1/2}$ corresponding to  $[9f]$ calculations, while the last row contains the experimental values.   }
\begin{ruledtabular}
\begin{tabular}{lrrrrrrrrrrrrr}
\\[-0.2cm]
                                                   
                            & \multicolumn{3}{c}{$^{4}\!D$} &&&  \multicolumn{3}{c}{$^{4}\!P$} &&&  \multicolumn{3}{c}{$^{2}\!P$}   \\
                                                \hline \\[-0.1cm]
 $( L_i S_i, L_j S_j)$        &  $orb$  &   $sd$  &  $orb + sd$ &&& $orb$ & $sd$  &  $orb + sd$   &&&   $orb$  & $sd$  &  $orb + sd$   \\ [0.1cm]
  \cline{2-4} \cline{7-9} \cline{12-14}\\[-0.2cm] 
                                                                                                                                                                                                                                                                      
($^{2}P$,$^{2}P$ )          & $-$ 479 & $-$64   & $-$543  &&& $-$ 876 & $-$117  & $-$ 993  &&& $-$1449  & $-$194   & $-$1643 \\            
($^{4}P$,$^{4}P$ )          &     164 &      11 &     175 &&&     958 &      64 &     1022 &&&      279 &       19 &      298 \\   
($^{4}D$,$^{4}D$ )          &    1002 &    1401 &    2403 &&&      05 &      07 &       12 &&&      400 &      559 &      959 \\
2$\times$($^{2}P$,$^{4}P$ ) &       0 & $-$19   & $-$19   &&&       0 &      61 &       61 &&&        0 & $-$43    &    $-$43 \\
2$\times$($^{2}P$,$^{4}D$ ) &       0 & $-$416  & $-$416  &&&       0 &   $-$41 & $-$41    &&&        0 &      457 &      457 \\
2$\times$($^{4}P$,$^{4}D$ ) &    2431 &     487 &    2918 &&&  $-$430 &  $-$ 86 & $-$516   &&&  $-$2003 & $-$401   &  $-$2404 \\
\\ 
 &$A^{orb}_{1/2}$& $A^{sd}_{1/2}$ & $A_{1/2}$  &&&$A^{orb}_{1/2}$ & $A^{sd}_{1/2}$& $A_{1/2}$&&&$A^{orb}_{1/2}$ & $A^{sd}_{1/2}$    & $A_{1/2}$  \\
 \\[-0.1cm] 
  \cline{2-4} \cline{7-9} \cline{12-14}\\[-0.2cm] 
  
              &    3118 &    1400 &    4518 &&&  $-$343 & $-$112 & $-$455  &&&  $-$2773 &      397 & $-$2376 \\
 \\
  MR-BP[9f]           &    3183 &    1499 &    4646$\dag$ &&&  $-$351 & $-$115 & $-$462$\dag$  &&&  $-$2824 &     467  & $-$2327$\dag$ \\
 \\
 
 Exp \cite {Huoetal:2018a}  &&&       $4541 \pm 50$       &&&&&        $-226 \pm 50$    &&&&&     $- 2378 \pm 80$           \\                                          
\end{tabular}
$\dag$ These totals  differ from $ orb + sd  $ because they include the contact contribution which is not zero in the  MR-BP[9f]  calculations (see text for more details). \\
\end{ruledtabular}
\end{table}

\begin{table}[H]
\caption{\label{tab:J32}Values of  $ A^{orb}_{3/2}(\gamma L_i S_i,\gamma L_j S_j)=orb$,  $ A^{sd}_{3/2}(\gamma L_i S_i,\gamma L_j S_j)=sd$, $ A_{3/2}(\gamma L_i S_i,\gamma L_j S_j)=orb + sd$ for  $^{4}\!D$ ,  $^{4}\!F$,  $^{4}\!P$   and $^{2}\!P$ states. At the bottom of the table we give the total values $A^{orb}_{3/2}$, $A^{sd}_{3/2}$ and  $A_{3/2}$ corresponding to  BP$[9f]$  calculations, while the last row contains the experimental values. }
\begin{ruledtabular}
\begin{tabular}{lrrrrrrrrrrrrrrrrrr}
\\[-0.2cm]                         
                                                     & \multicolumn{3}{c}{$^{4}\!D$}  &&& \multicolumn{3}{c}{$^{4}\!F$}  &&&   \multicolumn{3}{c}{$^{4}\!P$} &&& \multicolumn{3}{c}{$^{2}\!P$}\\ \hline \\[0.1cm]
$( L_i S_i, L_j S_j)$                  &  $orb$  &   $sd$  &  $orb + sd$   &&&   $orb$  & $sd$  &  $orb + sd$  &&& $orb$  &   $sd$  &  $orb + sd$ &&&   $orb$  &   $sd$  &  $orb + sd$ \\   
 [+0.2cm]                                                                                 \cline{2-4} \cline{7-9} \cline{12-14} \cline{17-19}\\[-0.2cm]                                                                                                                                       
($^{2}D$,$^{2}D$ )                   &     138     &    129    &      267     &&&      0     &       0   &        0          &&&      146     &      137       &      283 &&&      111    &     104    &      215  \\    
($^{2}P$,$^{2}P$ )                    & $-$97    &     3    &   $-$94      &&& $-$381 &      10   & $-$371       &&&    $-$78   &          2       & $-$76  &&& $-$675    &       18   & $-$657  \\        
($^{4}P$,$^{4}P$ )                    & $-$46    & $-$10  &  $-$56       &&& $-$19   &  $-$4   & $-$23          &&&    $-$335 &    $-$76      & $-$411 &&& $-$36     &  $-$8     & $-$44  \\   
($^{4}F$,$^{4}F$ )                    &  118      & $-$43  &        75       &&& 1559   & $-$561 &    998       &&&           0  &       0        &      0        &&&     552    & $-$199   &     353   \\ 
($^{4}D$,$^{4}D$ )                   &  355      &   498   &       853     &&&       0   &       0   &        0          &&&         97  &      136     &      233  &&&      43     &       61    &     104   \\
2$\times$($^{2}D$,$^{2}P$ )    &  179      & $-$72  &       107      &&& $-$18    &      7    & $-$11     &&&   $-$165  &         66      & $-$ 99   &&& $-$424   &    170    & $-$254  \\ 
2$\times$($^{2}D$,$^{4}P$ )    &    0       & $-$9  &   $-$9          &&&     0     &      0    &       0             &&&            0  &    $-$24     &    $-$24 &&&       0     &       7   &       7  \\ 
2$\times$($^{2}D$,$^{4}F$ )    &    0      &      105 &        105       &&&     0     &      19    &       19        &&&            0  &    4    &   4      &&&       0    & $-$ 203  & $-$203  \\
2$\times$($^{2}D$,$^{4}D$ )    &    0      & $-$127 & $-$127     &&&     0     &       0    &        0         &&&            0  &         68    &       68     &&&        0    &    $-$40 & $-$40  \\ 
2$\times$($^{2}P$,$^{4}P$ )     &    0      &        8  &         8       &&&     0      &      11    &       11      &&&           0   &    $-$20     & $-$20     &&&        0    &       19   &       19 \\ 
2$\times$($^{2}P$,$^{4}F$ )     &    0      &       34  &       34        &&&   0      & $-$245  & $-$245      &&&          0   &          $-$1      &        $-$1      &&&        0    &      194   &      194  \\
2$\times$($^{2}P$,$^{4}D$ )    &    0      & $-$106 & $-$106      &&&     0      & $-$3     & $-$3        &&&          0   &    $-$50      & $-$50      &&&        0    &      98  &      98  \\
2$\times$($^{4}P$,$^{4}D$ )    & 686      &        0  &       686        &&&       7   &         0   &       7    &&& $-$968   &        0        & $-$968 &&& $-$211  &     0       & $-$211  \\  
2$\times$($^{4}F$,$^{4}D$ )    & 342     &        0   &       342      &&& $-$18    &       0     &  $-$18     &&&          $-$7  &        0         &    $-$7       &&& $-$259   &    0      & $-$259  \\ 
2$\times$($^{4}P$,$^{4}F$ )     &    0     &       66  &           66    &&&       0    & $-$156  &  $-$156     &&&      0     & 7          &   7         &&&       0    &     126  &      126  \\ 
\\
&$A^{orb}_{3/2}$& $A^{sd}_{3/2}$ & $A_{3/2}$         &&&$A^{orb}_{3/2}$ & $A^{sd}_{3/2}$    & $A_{3/2}$   &&&  $A^{orb}_{3/2}$ & $A^{sd}_{3/2}$    & $A_{3/2}$ &&&     $A^{orb}_{3/2}$ & $A^{sd}_{3/2}$    & $A_{3/2}$ \\
\\
 \cline{2-4} \cline{7-9} \cline{12-14} \cline{17-19}\\[-0.2cm]
               &   1675  &      476  &      2151   &&&     1130  & $-$922  &      208     &&&  $-$1310  &    249    & $-$1061   &&& $-$899  &     347   & $-$552  \\
  \\
  MR-BP[9f]             &   1706  &      507  &   2262$\dag$      &&&     1146 & $-$1009 &       122$\dag$     &&&  $-$1333  & 281       & $-$1033$\dag$   &&& $-$916 &       382   & $-$496$\dag$\\  
  \\            
 Exp \cite {Huoetal:2018a}          &            &         & 2290$\pm$50 &&&&              &$110 \pm 10$ &&              &                   && $-1035 \pm 50$ &&&&              & $-498 \pm 80$            \\      
\end{tabular}
$ \dag$ These totals differ from $ orb + sd  $ because they include the contact contribution which is not zero in the  MR-BP[9f]  calculations (see text for more explanations). \\

\end{ruledtabular}
\end{table}  

\begin{table}[H]
\caption{\label{tab:J52}Values of  $ A^{orb}_{5/2}(\gamma L_i S_i,\gamma L_j S_j)=orb$,  $ A^{sd}_{5/2}(\gamma L_i S_i,\gamma L_j S_j)=sd$, $ A_{5/2}(\gamma L_i S_i,\gamma L_j S_j)=orb + sd$ for  $^{4}\!D$ ,  $^{4}\!F$,  $^{2}\!F$   and $^{4}\!P$ states. At the bottom of the table we give the total values $A^{orb}_{5/2}$, $A^{sd}_{5/2}$ and  $A_{5/2}$ corresponding to  BP$[9f]$  calculations, while the last row contains the experimental values.  }
\begin{ruledtabular}
\begin{tabular}{lrrrrrrrrrrrrrrrrrr}
\\[-0.2cm]
 $( L_i S_i, L_j S_j)$               &\multicolumn{3}{c}{$^{4}\!D$}&&& \multicolumn{3}{c}{$^{4}\!F$}&&&  \multicolumn{3}{c}{$^{2}\!F$}&&& \multicolumn{3}{c}{$^{4}\!P$}\\ \hline \\[0.1cm]   
                                                  &  $orb$  &   $sd$  &  $orb + sd$ &&&   $orb$  & $sd$  &  $orb + sd$  &&& $orb$  &   $sd$  &  $orb + sd$ &&&   $orb$  &   $sd$  &  $orb + sd$ \\   
 [+0.2cm]                                                                                                                                                                          \cline{2-4} \cline{7-9} \cline{12-14} \cline{17-19}\\[-0.2cm]                                             
($^{2}D$,$^{2}D$ )                &     46   &  $-$18   &        28       &&&      87 & $-$35  &       52         &&&     33  & $-$13   &       20           &&&       1   &   0   &       1            \\  
($^{2}F$,$^{2}F$ )                 &       11   & $-$3  &        8          &&&     531 & $-$128&      403        &&&    323  & $-$77 &      246          &&&     547 & $-$131 &      416         \\  
($^{4}P$,$^{4}P$ )                 & $-$53  &        2  &    $-$51      &&&       $-$1 & 0        &    $-$1          &&& $-$390&    16    & $-$374           &&& $-$327 &    13    & $-$314         \\  
($^{4}F$,$^{4}F$ )                 &    149  & $-$43  &    106          &&&   647  & $-$187 &     460        &&&   184  & $-$53   &      131             &&&     364 & $-$105  &    259            \\  
($^{4}D$,$^{4}D$ )                &    326  &     219 &      545        &&&      18 &     11     &       29         &&&       63 &       42 &      105           &&&       1   &       01   &       2           \\  
2$\times$($^{2}D$,$^{2}F$ ) &       15   &        23  &        38          &&&$-$135& $-$217 & $-$352       &&& $-$65  & $-$104 &  $-$169          &&&       11 &       17  &       28           \\  
2$\times$($^{2}D$,$^{4}P$ ) &       0   &       13 &       13       &&&       0 & $-$3     & $-$3            &&&       0  &       30  &       30              &&&       0   &       4   &       4             \\  
2$\times$($^{2}D$,$^{4}F$ ) &       0   &       56 &       56         &&&       0 & $-$160 & $-$160        &&&       0   &       52 &       52             &&&       0   & $-$9   &  $-$9          \\  
2$\times$($^{2}D$,$^{4}D$ ) &       0  & $-$126 & $-$126        &&&       0 & $-$40  & $-$40           &&&       0  &       46  &       46             &&&       0 &       1     &       1  \\  
2$\times$($^{2}F$,$^{4}P$ ) &       0   & $-$4   & $-$4             &&&      0   & $-$4     & $-$4             &&&       0 &       60  &       60             &&&       0 & $-$72    & $-$72  \\  
2$\times$($^{2}F$,$^{4}F$ ) &       0   &       8   &       8            &&&      0  &      120  &      120         &&&       0 & $-$50   & $-$50           &&&       0 & $-$91   & $-$91  \\  
2$\times$($^{2}F$,$^{4}D$ ) &       0  & $-$5    & $-$5             &&&      0  &        7  &        7            &&&       0 & $-$11 & $-$11           &&&       0 &       2    &       2  \\  
2$\times$($^{4}P$,$^{4}D$ ) &    320 & $-$107 &      213         &&& $-$10 &       3 & $-$7          &&&$-$380&      127 & $-$253      &&& $-$47 &       16 & $-$31  \\  
2$\times$($^{4}F$,$^{4}D$ ) &      326 &   163 &      489          &&& $-$158 & $-$79 & $-$237       &&& $-$159 & $-$80 & $-$239         &&&    31    &    15    &       46  \\  
2$\times$($^{4}P$,$^{4}F$ ) &       0    &    59   &    59         &&&      0    &       17 &       17        &&&      0    &      177 &  177               &&&       0 & $-$228  & $-$228  \\  
\\
      &$A^{orb}_{5/2}$& $A^{sd}_{5/2}$ & $A_{5/2}$ &&&$A^{orb}_{5/2}$ & $A^{sd}_{5/2}$    & $A_{5/2}$   &&&  $A^{orb}_{5/2}$ & $A^{sd}_{5/2}$    & $A_{5/2}$ &&&     $A^{orb}_{5/2}$ & $A^{sd}_{5/2}$    & $A_{5/2}$ \\ 
       \cline{2-4} \cline{7-9} \cline{12-14} \cline{17-19}\\[-0.2cm]
\\
             &     1140 &      237 &     1377               &&&    979 & $-$695 &      284       &&& $-$391 &  162    & $-$229               &&&    581 & $-$567 & 14  \\
 \\  
MR-BP[9f]              &   1157 &  253     &     1461$\dag$       &&&    990 & $-$764 &     252$\dag$      &&& $-$400 &    180 &   $-$202$\dag$    &&& 589 & $-$622 & $-$36$\dag$ \\          
\\        
Exp \cite {Huoetal:2018a}                                                           &&&           $1481 \pm 20$              &&&&&      $304 \pm 50$       &&&&&   $- 190 \pm 10$ &&&&&   $-17 \pm 10$ \\      
\end{tabular}
$\dag$ These totals differ from $ orb + sd  $ because they include the contact contribution which is not zero in the  MR-BP[9f]  calculations (see text for more explanations). \\
\end{ruledtabular}
\end{table}

\begin{table}[H]
\caption{\label{tab:J72} Values of  $ A^{orb}_{7/2}(\gamma L_i S_i,\gamma L_j S_j)=orb$,  $ A^{sd}_{7/2}(\gamma L_i S_i,\gamma L_j S_j)=sd$, $ A_{7/2}(\gamma L_i S_i,\gamma L_j S_j)=orb + sd$ for  $^{4}\!D$ ,  $^{4}\!F$ and $^{2}\!F$   states. At the bottom of the table we give the total values $A^{orb}_{7/2}$, $A^{sd}_{7/2}$ and  $A_{7/2}$ corresponding to  BP$[9f]$  calculations, while the last row contains the experimental values.   }
\begin{ruledtabular}
\begin{tabular}{lrrrrrrrrrrrrr}
\\[-0.2cm]

                            & \multicolumn{3}{c}{$^{4}\!D$} &&& \multicolumn{3}{c}{$^{4}\!F$}  &&& \multicolumn{3}{c}{$^{2}\!F$}     \\
                                                 \hline \\[0.1cm] 
$( L_i S_i, L_j S_j)$         &  $orb$  &   $sd$  &  $orb + sd$ &&& $orb$ & $sd$  &  $orb + sd$   &&&   $orb$  & $sd$  &  $orb + sd$   \\   
 [+0.2cm]                                                                                 \cline{2-4} \cline{7-9} \cline{12-14} \\[-0.2cm]                                                                                                                                       
($^{2}F$,$^{2}F$ )          &      46 &       6 &      52 &&&      257 &      34 &     291 &&&      901 &       120 &     1021 \\                     
($^{4}F$,$^{4}F$ )          &     127 &   $-$13 &     114 &&&      673 &   $-$67 &     606 &&&      270 &     $-$27 &      243 \\            
($^{4}D$,$^{4}D$ )          &     338 &  $-$203 &     135 &&&       63 &   $-$38 &      25 &&&        0 &         0 &        0 \\            
2$\times$($^{2}F$,$^{4}F$ ) &       0 &   $-$22 &   $-$22 &&&        0 &  $-$119 &  $-$119 &&&        0 &       141 &      141 \\            
2$\times$($^{2}F$,$^{4}D$ ) &       0 &      61 &      61 &&&        0 &   $-$62 &   $-$62 &&&        0 &         1 &        1 \\            
2$\times$($^{4}F$,$^{4}D$ ) &     207 &     249 &     456 &&&   $-$206 &  $-$248 &  $-$454 &&&     $-$1 &     $-$ 2 &    $-$ 3 \\            
\\
      &$A^{orb}_{7/2}$& $A^{sd}_{7/2}$ & $A_{7/2}$  &&&$A^{orb}_{7/2}$ & $A^{sd}_{7/2}$& $A_{7/2}$&&&$A^{orb}_{7/2}$ & $A^{sd}_{7/2}$    & $A_{7/2}$    \\ [-0.1cm] 
\\
   \cline{2-4} \cline{7-9} \cline{12-14} \\[-0.2cm]      
\\
      &     718 &      78 &     796 &&&      787 &  $-$500 &     287 &&&     1170 &       233 &     1403 \\
 \\
 MR-BP[9f]     &     729 &      82 &     852$\dag$ &&&     799 &   $-$553 &   263$\dag$   &&&     1187 &       247 &     1480$\dag$ \\
   \\  
 Exp \cite {Huoetal:2018a}  &&&             $793 \pm 20$  &&&&&                 $276 \pm 10$ &&&&&                                 \\           
\end{tabular}
$\dag$ These totals are slightly different from $ orb + sd  $ because they include the contact contribution which is not zero in the  MR-BP[9f]  calculations (see text for more explanations). \\
\end{ruledtabular}
\end{table}

\clearpage
 \section{Conclusion}
 \label{sec:concl}
In this work, we present the results of elaborate {\it ab initio} variational calculations of hyperfine constants for 17 levels in fluorine, all arising from the 6 terms 
$2p^4 (^3\!P)3d~\; ^4\!D, \; ^2\!D, \; ^4\!F, \; ^2\!F, \; ^4\!P$ and $^2\!P$.
 The choice of these levels was guided and justified by the recent publication of experimental $A_J$ values for 15 of these 17 levels, extracted  from  concentration modulation spectroscopy experiments~\cite{Huoetal:2018a}. The global theory-observation agreement is very  good ($\approx 3.5$\%) for 13 levels, taking into account of the relatively large experimental uncertainty of the order of 5\%.
  The larger disagreement observed for  $A(^4P_{1/2})$ and $A(^4P_{5/2})$ can be fully understood in terms of large cancellation and interference effects that make their estimation particularly challenging.

The present theoretical study is at first sight quite surprising, although some previous work on other levels of fluorine atom
opened this perspective~\cite{Caretal:2013b,Aouetal:2018a}. It indeed reveals, in contrast to what is a priori expected for light atoms,  weak electron correlation effects on hyperfine structures, but large (if not huge) relativistic effects on hyperfine constants.
 To explain the latter observation, we investigated the matrix elements of the magnetic dipole hyperfine interaction Hamiltonian  in the limited   $\{ 2p^4 (^3\!P) 3d \; L_iS_i\}$ configuration space, extracting the weights from the eigenvectors of much larger CSF expansions.
 This detailed analysis, combining the Breit-Pauli wave function compositions, with the analytical Racah algebra ingredients, beautifully illustrates the crucial role of relativistic  term-mixing in the theoretical estimation of the hyperfine constants. It also sheds invaluable light on the interference mechanism between the orbital and spin-dipole contributions, and between the relativistic coupling-term contributions to the hyperfine constant values, allowing to understand their relative magnitude. 
 
 Estimations and investigations of theoretical uncertainties of atomic properties should be systematically included, when possible. As observed by Drake~\cite{Dra:2020a}, it is clear that the culture is changing within the theoretical computational community to make uncertainty quantification (UQ) the usual expectation when theoretical results are presented. The present work is one step in this direction, as a  few others in the framework of multiconfiguration variational approaches~\cite{Papetal:2019a,Gaietal:2020a,Papetal:2021a}.  It indeed illustrates how the details of the magnetic dipole hyperfine operators can be explored to point difficult cases in terms of cancellation, either between $LS$ pairs for individual operators, or between the orbital and the spin-dipolar operators,  and to asses the reliability of the theoretical hyperfine constants. 
As an example, the relative large uncertainty  inferred from the observed differences between MCHF-BP and MCDHF-RCI, as well as from the theory-experiment differences  for the two levels $2p^4(^3\!P)3d \; ^4\!P_{1/2}$ and $^4\!P_{5/2}$  can be explained by large interferences occurring in the amplitude of the observable.

Incidentally, the perfect consistency between the Breit-Pauli calculations and the RCI-P approaches was demonstrated. Orbital orthogonality constraints in the BP calculations forcing the use of a simultaneous optimization strategy in the MCHF approach, and the  layer-by-layer approach used to solve convergence issues in the fully relativistic MCDHF scheme \cite{Schetal:2020a}, are the current limiting factors to guarantee the consistency between the two approaches. The global agreement  between the two methods is however good for similar configuration lists and orbital active sets used to build the variational spaces.

\begin{acknowledgments}
F.Z.B. and M.N. acknowledge financial support from the Direction G\'en\'erale de la Recherche Scientifique et du D\'evelopement Technologique (DGRSDT) of Algeria.  
M.G. acknowledges support from the FWO \& FNRS Excellence of Science Programme (EOS-O022818F).
P.J.  acknowledges support from the
Swedish research council under contract
and 2016-04185.
\end{acknowledgments}


\end{document}